# Mortality data reliability in an internal model

Fabrice Balland[1], Alexandre Boumezoued[2], Laurent Devineau[2], Marine Habart[1], Tom Popa[1]

**Abstract**

In this paper, we discuss the impact of some mortality data anomalies on an internal model capturing longevity risk in the Solvency 2 framework. In particular, we are concerned with abnormal cohort effects such as those for generations 1919 and 1920, for which the period tables provided by the Human Mortality Database show particularly low and high mortality rates respectively. To provide corrected tables for the three countries of interest here (France, Italy and West Germany), we use the approach developed by Boumezoued (2016) for countries for which the method applies (France and Italy), and provide an extension of the method for West Germany as monthly fertility histories are not sufficient to cover the generations of interest. These mortality tables are crucial inputs to stochastic mortality models forecasting future scenarios, from which the extreme 0,5% longevity improvement can be extracted, allowing for the calculation of the Solvency Capital Requirement (SCR). More precisely, to assess the impact of such anomalies in the Solvency II framework, we use a simplified internal model based on three usual stochastic models to project mortality rates in the future combined with a closure table methodology for older ages. Correcting this bias obviously improves the data quality of the mortality inputs, which is of paramount importance today, and slightly decreases the capital requirement. Overall, the longevity risk assessment remains stable, as well as the selection of the stochastic mortality model. As a collateral gain of this data quality improvement, the more regular estimated parameters allow for new insights and a refined assessment regarding longevity risk.

---

[1] GIE AXA, 21, Avenue Matignon, 75008 Paris, France

[2] Milliman, 14 Avenue de la Grande Armée, 75017 Paris, France

# Contents

# 1. Introduction

In this paper, our aim is to discuss and assess the impact of anomalies in national mortality tables provided by the Human Mortality Database[3] on the internal model of a typical insurer which captures longevity risk in the Solvency 2 framework. Of special interest in this paper are some abnormal cohort effects observed on period tables in which some diagonals show special patterns. Particular attention is devoted to the 1919-1920 effect: the 1919 diagonal shows particularly low mortality rates, whereas the 1920 generation shows particularly high ones. This effect can be easily observed using matrices of mortality improvements by age and time, on which clear diagonal effects appear, or by comparing the mortality rates of the generations considered, see the example of France in Figure 1.

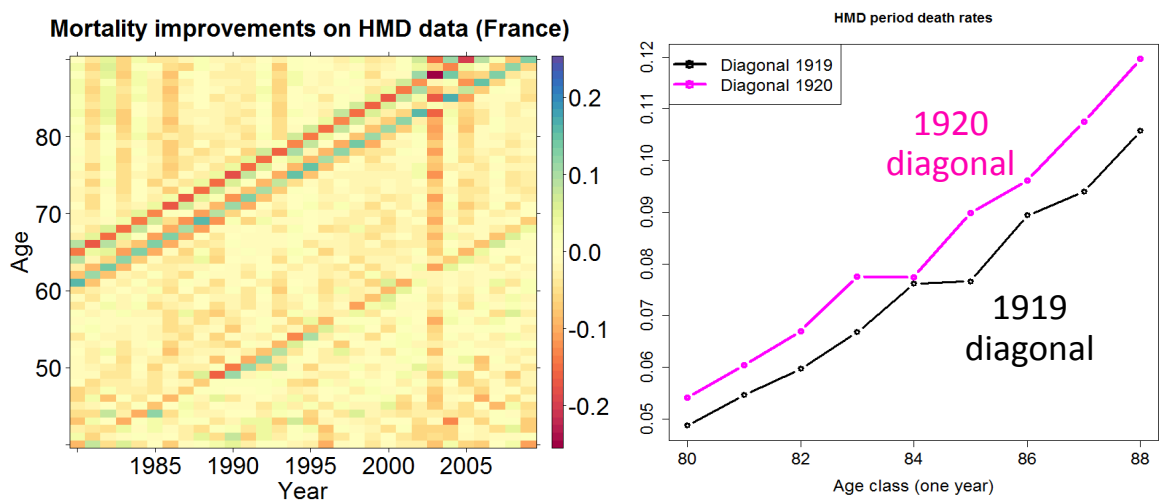

Figure 1 - Abnormal cohort effects: example of the period table for France

Such results are counter-intuitive regarding natural demographic insight: some populations born more recently live less longer in average (see again Figure 1), as it is indeed observed that the "younger" generation 1920 shows higher mortality rates in the period table.

In the actuarial literature, few references have been focusing on mortality data reliability. The awareness about such anomalies has recently emerged from the work by Richards (2008), Cairns et al. (2016) and Boumezoued (2016) (see also Boumezoued & Devineau (2017)). To our knowledge, the first conjecture about the potential causes of these isolated cohort effects was by Richards (2008). He focused on the 1919 birth cohort for England & Wales for which he suggested the possibility of errors due to erratic number of births. Such conjecture took a concrete form as the ONS (Office for National Statistics) produced corrected tables, in fact a mortality increase for this 1919 cohort, particularly at high ages. The ONS methodology has then been studied by Cairns et al. (2016) in several directions, who proposed an approach whose scope was to correct the mortality table of England & Wales. Then, Boumezoued (2016) underlined the universality of such anomalies (isolated cohort effects) by highlighting them in a variety of countries from the Human Mortality Database (HMD). By studying the HMD methodology to construct period tables, he proposed to source its fertility counterpart, the Human

---

[3] *Human Mortality Database. University of California, Berkeley (USA), and Max Planck Institute for Demographic Research (Germany). Available at www.mortality.org or www.humanmortality.de*. It is worth mentioning that at the time of writing, the Human Mortality Database released an update on February 2018, including in particular a revision of exposure calculation based on monthly birth counts. We refer the reader to the Version 6 of the HMD Methods Protocol for more details.

Fertility Database (HFD), and to apply the philosophy of the work by Cairns et al. (2016) to produce corrected period mortality tables for a set of countries.

In this paper, we discuss the impact of some mortality data anomalies on the longevity risk module of an internal model in the framework of Solvency 2, and in particular on trend risk which can be defined as the risk that the trend driving the future longevity evolution may experience unexpected changes. The three countries of interest in this paper are France, Italy and West Germany, due to their business exposure, and as they have been identified as embedding anomalies in their HMD mortality data. By applying and extending the correction methodology as described in Boumezoued (2016), we are able to measure the discrepancy between original and corrected period mortality tables. These discrepancies are crucial in the way the insurance market measures and manages longevity risk, especially in the present context.

Among life risks modules, one of the most important risks threatening insurers is longevity risk, which is the risk that insured people may on average survive longer than expected. The longevity exposure is the result of long term commitments (up to 60 years) with high uncertainties in the mid-term (10-20 years). Several decades and therefore several generations of contracts might be necessary before detecting any risk deviation. There is no consensus on the future evolution of longevity, either in terms of model or market price. Many interconnected factors are to be considered in the future longevity trend, as for example technological improvements, socio-economic trends, political systems, demographic structures. While some experts consider that the human being has already reached its maximum age, others dream of immortality thanks to transhumanism. Facing with such uncertainty, preparing insurance companies for aging issues and its business consequences is fundamental.

As a reminder, the Solvency II Directive became fully applicable on 2016, January 1$^{st}$. Solvency II reviews the prudential regime for insurance and reinsurance undertakings in the European Union in a harmonized prudential framework. The risk profile of each individual insurance company is further considered, with the aim to promote comparability, transparency and competitiveness. The Solvency II Directive provides two ways of measuring risk: insurance and reinsurance companies can use either the standard formula or their own internal model, which enables them to assess their own risk more accurately. In the framework of the Standard Formula, a risk classification is provided, which includes risk modules (such as market risks, life risks, non-life risks) and risk sub modules (such as mortality, longevity, lapse, CAT, etc. for life risks module). For each sub module, a stress test or a closed formula is used to determine a capital charge. The capital requirement is then calculated using a bottom-up approach, by aggregating submodules and then modules based on a correlation matrix. In comparison, the development of an internal model allows the company to refine its own risk assessment. In particular regarding life risks, the company has the possibility to use external data which reflects its risk in terms of countries considered, such as general population tables as provided by the Human Mortality Database for more than thirty countries and regions worldwide. It is the purpose of the present paper to identify, correct and test the impact of anomalies in national general population tables as key inputs in the internal model.

It is worth mentioning that the longevity shocks calibration for the Standard Formula has resulted from the comparison of retrospective and prospective analyses, see the CEIOPS' Advice for Level 2 Implementing Measures on Solvency II, especially the Annex B "Longevity risk calibration analysis". More specifically, two analyses have been conducted: the first one focusing on past historical mortality improvements, the second one generating shocks in the future. It has been concluded that the direct use of historical data led to a higher longevity stress compared to that derived from the forecasts of a stochastic model. In the same spirit, a prospective and retrospective analysis has been derived by Boumezoued (2016) on both crude and corrected mortality data; the analysis of corrected mortality tables showed that, not only the mortality levels for several cohorts were highly over/under-estimated, but also that the volatility of mortality improvement rates over the last 30 years was over-estimated in original HMD mortality tables, with a large difference for many countries. In addition, it has been shown how the volatility levels reproduced by classical stochastic mortality models now closely match historical mortality improvements in corrected tables, although it was not the case on crude data as pointed out by several studies as cited before.

The previous considerations illustrate the potential impact of such mortality data correction on several steps of internal modelling process. To assess the business impact of such a complex risk, we use in this paper a simplified internal model specification. Market practices rely on two main classes of stochastic mortality models: the first class regroups models derived from Lee-Carter approach while the second class of models is derived from Cairns, Black and Dowd approach. Usually calibrated on national mortality tables, such models have proven their ability to capture historical mortality behaviors. However, a core issue is their sensitivity to possible singularities or anomalies in the underlying mortality data.

In this paper, a simplified prototype of internal model is used and embeds the following features: a set of mortality models is fitted based on national population mortality data, and the 'best' model is selected based on statistical and qualitative criteria. The selected mortality model is then used to draw simulated forecasts, and a prospective mortality table corresponding to the 99.5 percentile in terms of improvements is built. This allows to provide the longevity shock at the core of the Solvency II economic capital calculation, here based on an annuity product portfolio.

The paper is organized as follows. Section 2 deals with the mortality data anomalies and focuses on the methodologies to detect and correct them, as well as on the key characteristics of corrected mortality tables. This includes the use of the method developed by Boumezoued (2016) to correct period tables for Italy and France, and the novel approach we propose in order to correct the West Germany mortality table, for which fertility histories are not sufficient to cover the 1919-1920 generations of interest. In Section 3, we introduce the stylized internal model which is used to calculate the Solvency Capital Requirement associated with longevity risk, and the impact of the mortality data anomalies on the internal model in the light of several analyses; in particular, we discuss the way it impacts the fit of stochastic mortality models, the future forecasts, the model selection process, the calculation of the SCR, as well as the stability of longevity risk assessment. The paper ends with some concluding remarks in Section 4.

## 2. Mortality data reliability issues

In this section, we focus on the identification and correction of several anomalies in mortality tables obtained from the Human Mortality Database for our three countries of interest: France, Italy and West Germany. After highlighting the recent awareness on such anomalies in the literature a well as the key observations on the crude datasets, we detail the correction methodologies we use and we describe the main characteristics of corrected mortality tables.

### 2.1. On the recent awareness about anomalies in mortality tables

As mentioned in the Introduction, few actuarial work has been focusing on mortality data reliability, but the awareness about such anomalies have recently emerged from the work by Richards (2008), Cairns et al. (2016) and Boumezoued (2016). The last author put into evidence the universality of such anomalies (isolated cohort effects) by highlighting them in a variety of countries from the Human Mortality Database (HMD). By studying the HMD methodology to construct period tables, he proposed to source its fertility counterpart, the Human Fertility Database (HFD), and to apply the philosophy of the work by Cairns et al. (2016) to produce corrected period mortality tables for a set of countries.

In this paper, we apply and extend the correction methodology as described in Boumezoued (2016). This way, we are able to measure the discrepancy between original and corrected period mortality tables, which are crucial inputs for the measurement and management of longevity risk in the insurance market.

In order to present the up-to-date awareness on such abnormal cohort effects, we summarize below the main conclusions from the work of Boumezoued (2016):

1. While comparing period and cohort mortality tables, anomalies in period ones have been highlighted in the form of isolated cohort effects for several countries available in HMD: the period mortality rates for specific generations appear surprisingly low or high compared to the others.

2. The HMD methodology to construct mortality estimates has been identified to embed a strong assumption of uniform distribution of births that is specific to the computation of period mortality tables, which shows that we are facing a universal reliability issue which is shared by most countries.

3. To perform an automatic correction procedure, it is proposed to rely on the Human Fertility Database (HFD), which is considered as the perfect counterpart of the HMD in terms of fertility.

4. The analysis of corrected mortality tables for several countries led to the conclusion that isolated cohort effects" in original mortality tables (often for years of birth around 1915, 1920, 1940 and 1945) are in fact universal anomalies that disappear in corrected tables. Further analysis of corrected mortality tables show that, not only the mortality levels for several cohorts are highly over/under-estimated, but also that the volatility of mortality improvement rates over the last 30 years was over-estimated in original HMD mortality tables, with a large difference for many countries.

In the next part 2.2, we detail the correction methodology based on fertility data which is used in the present paper to correct mortality tables for France and Italy. In the subsequent part 2.3, we motivate the need for an extension of the methodology and we propose a novel approach to produce a corrected table for West Germany. Finally, subsection 2.4 details the main characteristics of the corrected mortality tables and the key discrepancies with the original data.

As a final remark, let us emphasize that two kinds of mortality tables are provided by the Human Mortality Database: *period* and *cohort*. Anomalies in terms of abnormal cohort effects have been identified in *period* mortality tables, which are naturally designed to study the dynamics of mortality from one year to the next, and therefore systematically used in practice to calibrate stochastic mortality models. For a review of the definition and construction methods for period and cohort tables, see Boumezoued (2016).

## 2.2. The correction method based on fertility data

In practice, period mortality tables for one-year time periods and one-year age classes are produced based on annual population estimates derived from census, as well as number of deaths combining the information of death reports and population counts. Note that in most countries, census may be performed out of the beginning of the year, or at intervals greater than one year, which leads the HMD to perform several adjustments, see Wilmoth et al. (2007) for more details. In our framework however, we assume that input data provided by the HMD as annual population counts and number of deaths in Lexis triangles are accurate.

Two ingredients are at the core of death rate computation for a given period: the number of deaths in this period, which we assumed to be reliable, divided by the so-called "exposure-to-risk" which represents the "quantity" of individuals at risk of death. In so-called period mortality tables, the exposure-to-risk takes the mathematical form of an integral over a one-year time period and a one-year age-class, see Boumezoued (2016) for more details on the formalism. In this setting the period mortality rate writes

$$m(x,t) = \frac{D(x,t)}{E(x,t)},$$

where $D(x,t)$ is the number of deaths and $E(x,t)$ the total time lived in the year $[t, t+1)$ at age between $x$ and $x+1$. The mortality rate is the core object to be modelled by stochastic mortality models, as it is also the case for the death probability $q(x,t)$, more interpretable as the probability for individuals aged $x$ at time $t$ to die in the next year, and linked to the mortality rates through the following equation:

$$q(x,t) = 1 - \exp(-m(x,t)).$$

In the standard computation practice, the exposure-to-risk component $E(x,t)$ is usually approximated based on annual population estimates, as it is done by reference providers as national institutes or the Human Mortality Database. In other words, the standard practice works with the average between the population at the beginning of the year and that at the end of the year, which is taken as a proxy for the integral over the year in continuous time. This computation amounts to assume that births are uniformly distributed in the year, as well as from one cohort to the next.

Although for several years where population flows are quite stable the assumption may seem reasonable (although not perfect), the approximation error can be huge in situations in which population numbers are fluctuating in the year, that is particularly the case when births are erratic due to demographic shocks (e.g. before and after wars and pandemics). Such specific patterns create a "convexity effect", and in this context it is no longer possible to approximate the exposure-to-risk by a standard average.

To correct for such approximation errors for countries available in the Human Mortality Database, the approach proposed by Boumezoued (2016) relies on monthly fertility records available from the Human Fertility Database. Such monthly population estimates are used to construct a "data quality indicator" assessing the reliability of estimates in the (period) mortality table for each generation. The indicator takes the form of a ratio which measures the deviation between an annual and a monthly approximation of the (annual) exposure-to risk. For each year of birth, this ratio is then used to adjust the estimates along each diagonal of the mortality table (that is following each cohort), producing a corrected table which does not present the initial anomalies.

In this latter work, corrected period tables have been provided for France, Switzerland, Finland, Sweden and Austria, countries for which complete and deep fertility histories were available at time of publication. Meanwhile, the monthly fertility data for Italy has been officially released by the Human Fertility Database, allowing us to use the same methodology for this country as well. Note also that this is the case of Iceland, which will be added to the set of explanatory countries for West Germany, see the next subsection c). In the following, we present the results in terms of correction indicator for France and Italy; the specific treatment of West Germany, for which the monthly fertility history is not deep enough, is detailed in the next part c). The key characteristics of corrected period tables are then described in the last part d) of this section.

The number of births by months as well as the correction indicator computed for France and Italy using the method by Boumezoued (2016) are depicted in Figure 2. This figure shows how shocks in birth patterns impact the exposure-to-risk approximation as it is performed in period tables provided by the HMD. The virtue of the correction indicator appears from these graphs: for a given year of birth, a value greater (resp. lower) than one indicates that the HMD period mortality rates for the generation are over-estimated (resp. under-estimated). In addition, the ratio measures the magnitude of the discrepancy between the annual exposure-to-risk approximation and its (better) monthly counterpart. It should be noted that although the dynamics of the number of births over the months differs between France and Italy, the 1919-1920 effect shows some universality: the period mortality rates for generation 1919 are under-estimated, whereas those for generation 1920 are over-estimated. Moreover, in each case the order of magnitude of the error is +/-6%, see again Figure 2; this may represent a level of particular attention for practitioners using national mortality data.

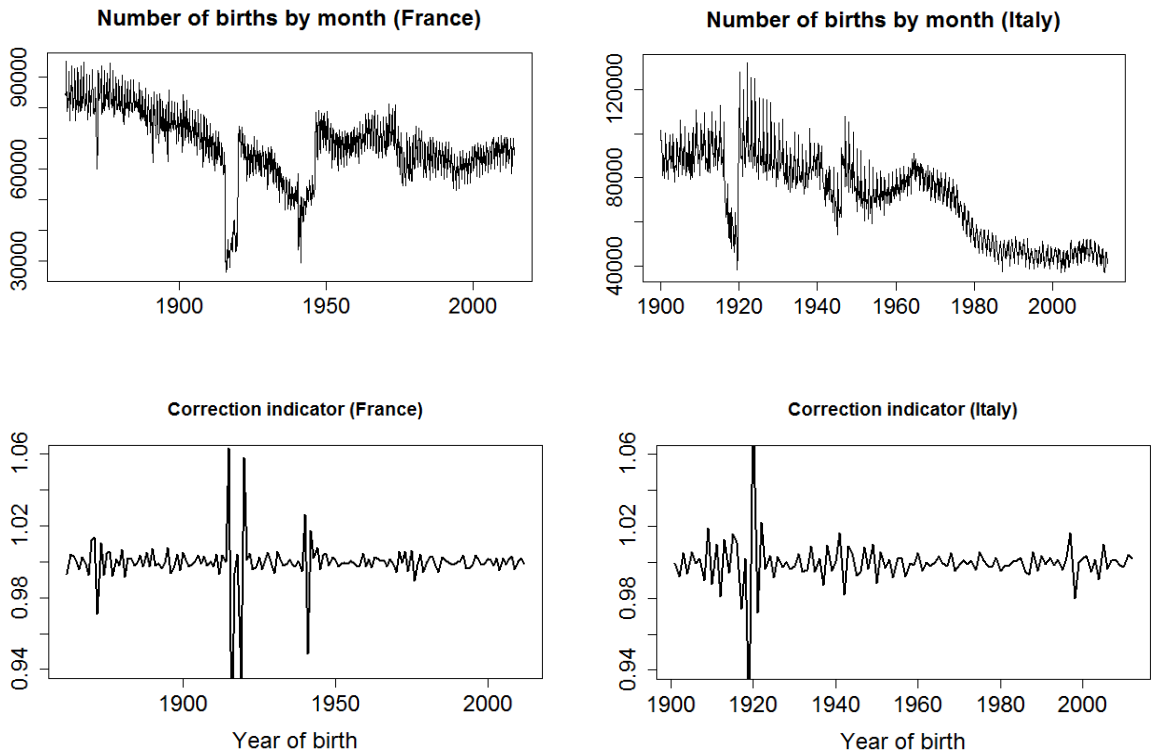

Figure 2 - Number of births by months and correction indicator for France and Italy

## 2.3. An extension of the correction method for West Germany

As mentioned in the previous part, in its present form the correction methodology can only be applied to years of birth for which the number of births by months are provided by the Human Fertility Database. Unfortunately, this is not the case for West Germany for which the HFD monthly fertility records are provided starting at year 1946, which consequently limits the set of years of birth for the correction ratio, see Figure 3.

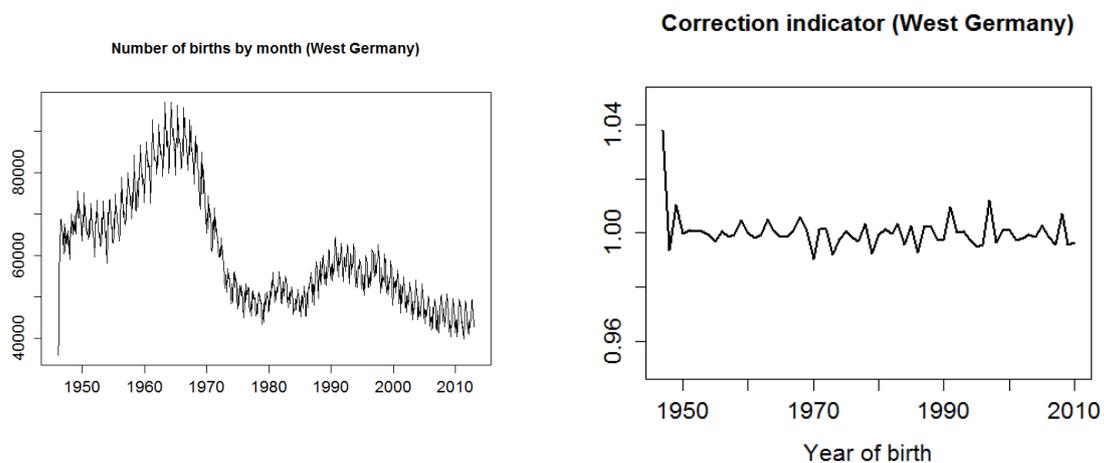

Figure 3 - Number of births by months and correction indicator for West Germany

Therefore, there is a need to extend the previous methodology to enable producing an output corrected table, especially for the typical 1919-1920 phenomenon which, again, lies outside of the historical period.

**Regression with stepwise selection process**

The idea now is to try to reconstruct the fertility history for West Germany, by looking at that other countries for which the birth series are available. Although number of births show clear different patterns from one country to another, (see Figure 2 in the previous part b), as well as Figure 9 in Boumezoued (2016)) the correction indicator, as a quantity without dimension (ratio of population estimates), shows quite similar patterns among the several countries.

To infer the correction indicator for West Germany starting from that of other countries, we propose to use a multiple regression approach, while performing an optimal selection procedure. In order to properly detail the regression and prediction methodology, let us introduce some notations. We denote by $I_C(t)$ the correction indicator available for country $C$ and year-of-birth $t$. The set of countries for which reasonable monthly fertility series are now available is set as

$$S = \{Finland, France, Iceland, Italy, Sweden, Switzerland\}.$$

This set is selected in order to satisfy the two following criteria:

- Data on births by months is officially released by the HFD (i.e. not stated as "preliminary release"), to avoid any major data issue on birth series,
- The monthly fertility series allow constructing correction indicators for years of birth before 1914, which covers the 1919-1920 effect with margin.

The aim now is to find an optimal set of regressors $S^* \subset S$, and coefficients $\{\alpha_C, C \in S^*\}$ and $\mu$ such that

$$I_{West\ Germany}(t) = \mu + \sum_{C \in S^*} \alpha_C\ I_C(t),$$

for each $t$ such as the indicator for West Germany $I_{West\ Germany}(t)$ can be constructed, that is here for each year of birth $t$ between 1947 and 2010 (2010 being chosen as an upper bound for all countries in order to avoid depending on recent revisions of demographic data).

**Results of the regression and prediction of the correction indicator**

The method carried out is a step-by-step comparison of the models obtained as a combination of regressors from the set $S$. At this stage, a statistical criterion has to be specified in order to compare models; we tested the classical criteria BIC and Adjusted R-square, both providing some penalization of the fitting ability of the model by the number of parameters. The results in terms of "best" model selected and associated parameter values are presented in the table below (rounded at 2 significant numbers).

|  | BIC | Adjusted R-squared |
| --- | --- | --- |
| $S^*$ | $\{Finland, France\}$ | $\{Finland, France, Italy\}$ |
| $\hat{\mu}$ | -0.19 | -0.14 |
| $\hat{\alpha}_{Finland}$ | 0.38 | 0.21 |
| $\hat{\alpha}_{France}$ | 0.81 | 0.63 |
| $\hat{\alpha}_{Iceland}$ | X | X |
| $\hat{\alpha}_{Italy}$ | X | 0.29 |
| $\hat{\alpha}_{Sweden}$ | X | X |
| $\hat{\alpha}_{Switzerland}$ | X | X |

Interestingly, the results show strong similarities between the BIC and the R-squared criteria. In particular, in each case the "best" model includes Finland and France as predictors of West Germany, although the Adjusted R-squared criterion also includes Italy. Another interesting feature is that the three countries of interest here, which have been identified to embed strong 1919-1920 anomalies, are linked with each other (with the Adjusted R-squared criterion), France and Italy being key predictors for West Germany. In terms of parameters, significant weight is given to France correction ratio as an explanatory variable in each case.

At this stage, empirical expert judgement has to be used to define the final choice between both criteria; in this paper, we choose the results given by the R-squared adjustments having in mind the three following arguments:

- It is preferred to use more series to increase stability, therefore to include Italy as well, as we consider it as a good input to explain fertility shocks in West Germany,
- The output corrected table for West Germany according to the R-squared criterion, see the next part d), shows graphically even less isolated cohort trends than that produced according to the BIC criterion,
- On the whole, both criteria (BIC or R-squared) provide similar results in terms of residuals, correction indicator series and output corrected tables.

**Prediction of the correction indicator**

For birth years outside the range of the correction indicator availability for West Germany, that is for $t$ before 1946, the *true* indicator $I(t)$ can be predicted by its estimate $\hat{I}(t)$ according to the regression formula

$$\hat{I}^{West\ Germany}(t) = \hat{\mu} + \sum_{C \in S^*} \hat{\alpha}_C\, I_C(t)$$

The reconstructed correction indicator is depicted in Figure 4. The key fact which appears from the observation of the reconstructed correction ratio is the following: although almost no spikes were observed in the available period for regression (plain line), the use of external data allows us to reproduce the generational fertility shocks at the core times in history (including around 1915 and around 1920) – this good feature captured here lies in the universality of the abnormal cohort countries in the selected regressors.

The correction efficiency of the regression approach presented here will be evaluated in the next part d) in which we analyze the main characteristics of the corrected mortality tables produced.

*Remark: It is worth mentioning that although this extrapolation over another time period may induce some instability issues in several regression contexts due to non-stationarity effects, we rely here on the fact that the selected regressors are 'mirror' countries which show similar error patterns (due to external and internal shocks), which are thus likely to reproduce the 1919-1920 effect (in between the Spanish flu and the end of the First World War). The illustrative results detailed in the next part indeed show the efficiency of the method to correct the 1919-1920 effects, while providing an underlying interpretation of the correction indicator dynamics based on 'explanatory' countries. At this stage, we argue that significant theoretical work remains to be done in this direction, which is out of scope of the present paper.*

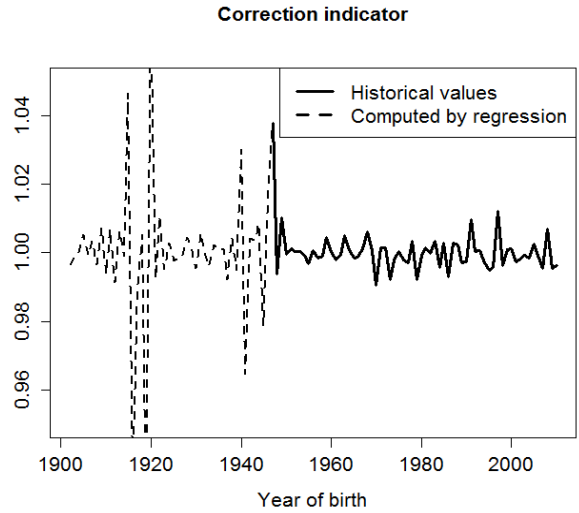

Figure 4 - Completion by regression of the correction indicator for West Germany

## 2.4. Analysis of corrected mortality tables

Let us denote by $I(b)$ the correction indicator for year of birth $b$. Then the corrected period mortality rates are constructed as

$$\widetilde{m}(x,t) = \frac{m(x,t)}{I(t-x)}$$

Then matrices of crude and corrected mortality improvement rates $r(x,t)$ or $\tilde{r}(x,t)$ can be analyzed and compared; mortality improvement rates are here defined as

$$r(x,t) = \frac{m(x,t+1)}{m(x,t)} - 1 \text{ and } \tilde{r}(x,t) = \frac{\widetilde{m}(x,t+1)}{\widetilde{m}(x,t)} - 1.$$

The data used to specify the crude death rates $m(x,t)$ by single year and one-year age bands were downloaded from the Human Mortality Database on September 1st 2015 for France (civil population), Italy (civil population) and West Germany (total population).

**Corrected mortality tables for France and Italy**

The matrices of crude and corrected mortality improvement rates by age and time are depicted in Figure 5. The matrices are centred on the 1919-1920 effects of interest, and the colour scale is fixed. The correction of these effects for France and Italy, and also the correction for France of the additional anomalies around years of birth 1915 and 1940 can clearly be observed. Note that for all the three countries, we use both male and female corrected tables, although we represent that related to the total population as a purpose of illustration.

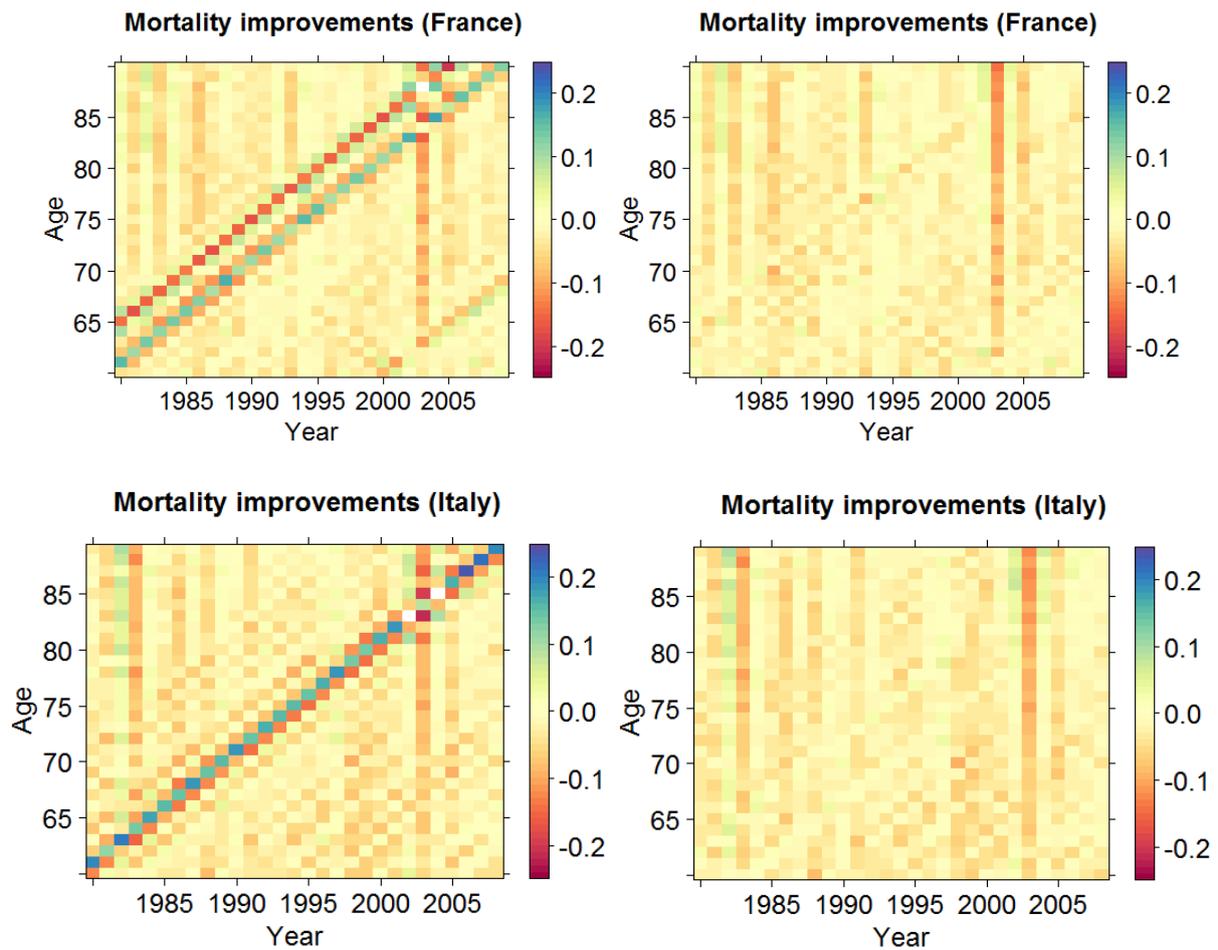

Figure 5 - Crude and corrected mortality improvements for France and Italy

**Corrected mortality tables for West Germany**

As stated in the previous part 2.3, the correction efficiency of the regression approach to reconstruct the correction indicator for West Germany is to be assessed by comparing the crude and corrected period mortality table. The results of the correction process are depicted on improvement rate matrices in Figure 6 for the two criteria selected (BIC or R-squared); from left to right: crude data, corrected data with BIC selection, corrected data with Adjusted R-squared selection.

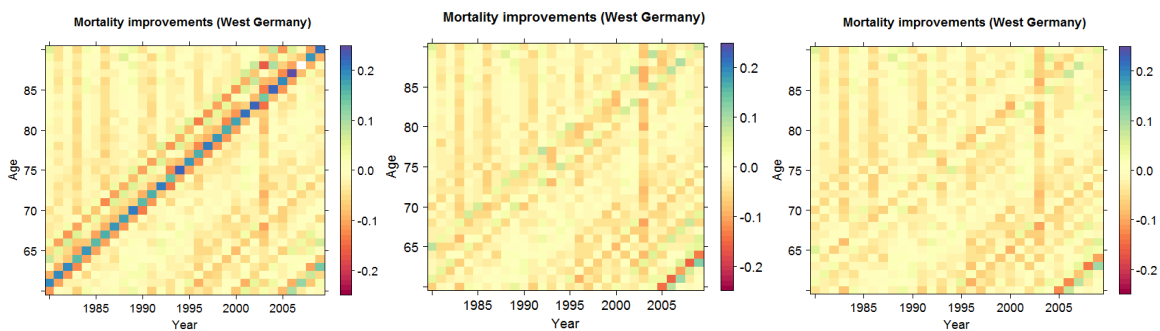

Figure 6 - From left to right: crude mortality improvements for West Germany, and the corrected versions according to the BIC and R-squared criteria

The correction process appears to be efficient concerning the 1919-1920 effects, which are the generations of interest in this paper. It should be noted that the 1915 effect is also clearly corrected, but additionally that the correction specific to year of birth 1945 correction is not fully performed. Although in our context, this is a second order issue as our main interest lies in the proper correction of the huge 1915 and 1920 anomalies, developing an improved correction method for such generations is left for further research.

Recall that the retained criterion is R-squared for the reasons detailed in the previous part 2.3, including in particular the fact that the corrected table shows even less isolated cohort trends.

**Key characteristics of corrected historical mortality tables**

To focus on historical data without introducing any external effect, analyses of the following section are conducted on mortality data embedded in the historical data tables only. As such, life expectancy and mortality indicators are restricted to ages 30 and 95.

While the method is applied on the entire mortality table, only specific generations are corrected and the main table characteristics are preserved, as shown for the period life expectancy of each year in Figure 7.

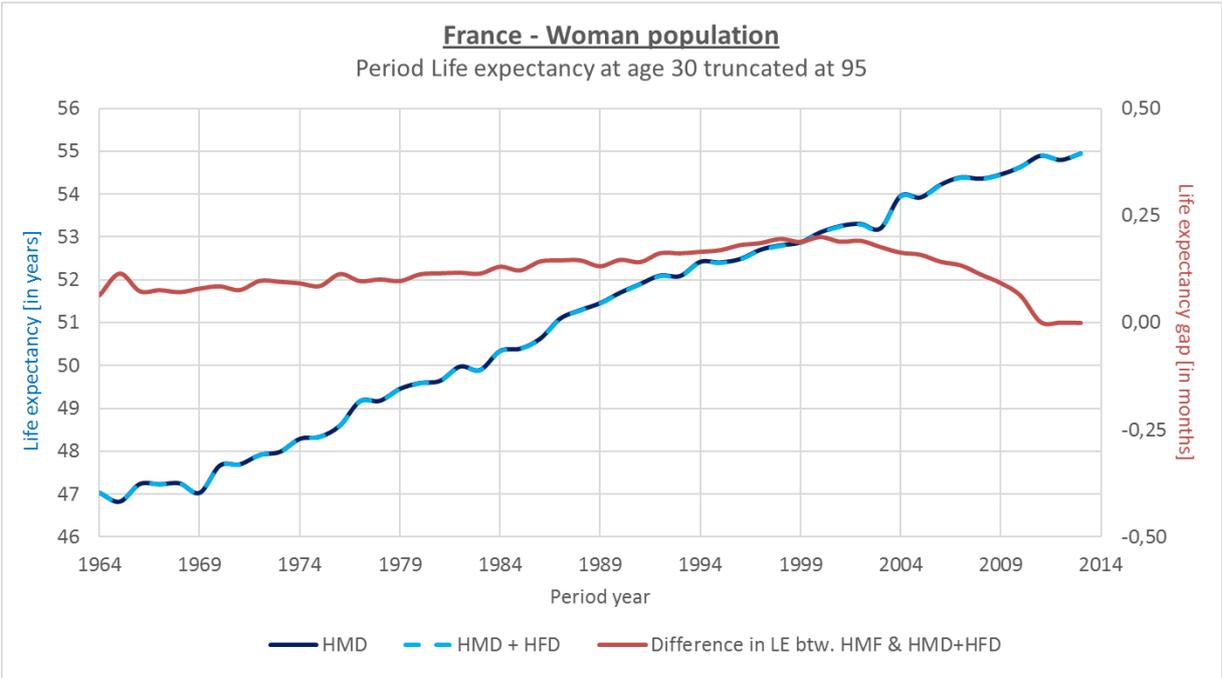

Figure 7 – Historical period life expectancy at age 30, truncated at age 95, before and after HFD correction

The following section highlights that these artificial cohort effects are identifiable on basic quantities usually derived on crude mortality data, as the force of mortality and the death functions.

**Period-based Forces of Mortality**

The analysis of the forces mortality of French woman reveals the transitory and specific volatility of generations born between 1918 to 1923 in the period based mortality tables of year 1964 to 2013.

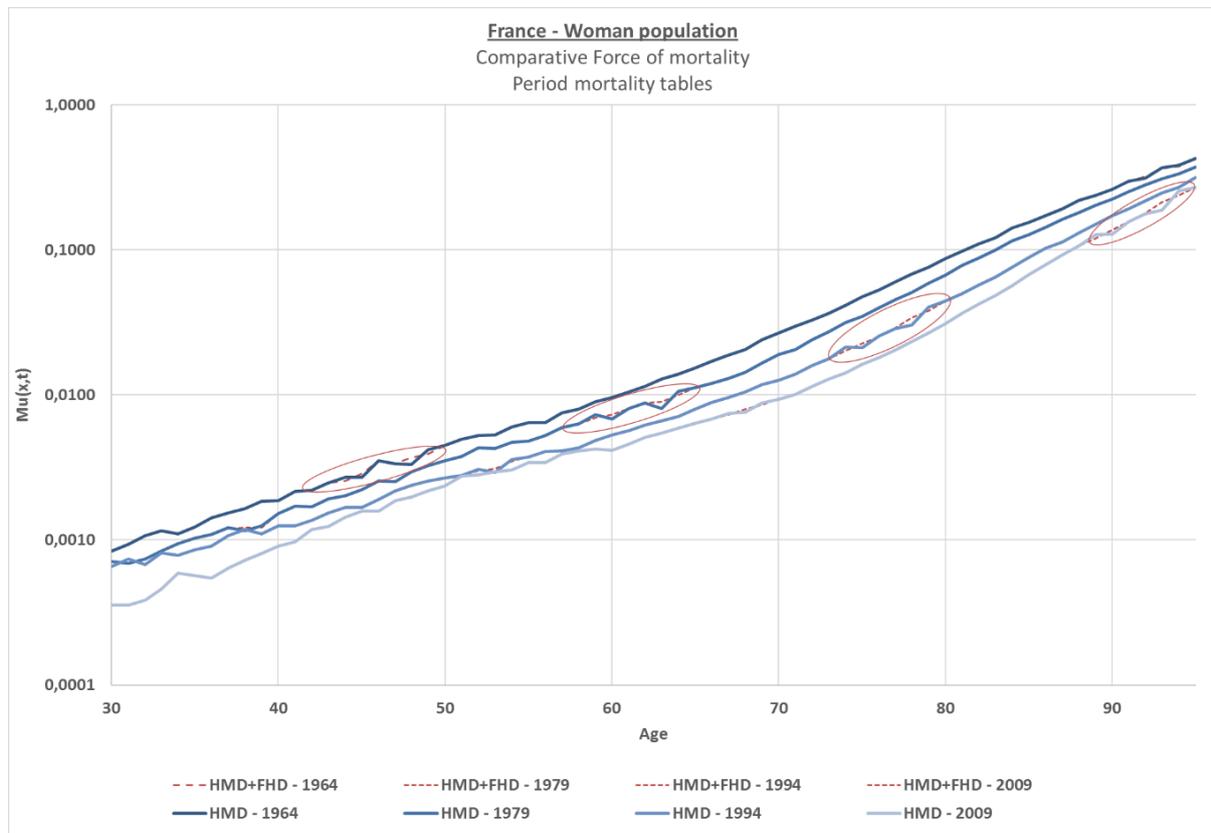

Figure 8 – Period force of mortality before and after HFD correction

**Death curves**

In mirror to the force of mortality, the next two graphs on Figure 9 present the death curves of France female population over the same 4 periods between years 1964 and 2009. It is noticeable how corrected tables increased the regularity of the death curves. Results are similar for all countries and gender in our study.

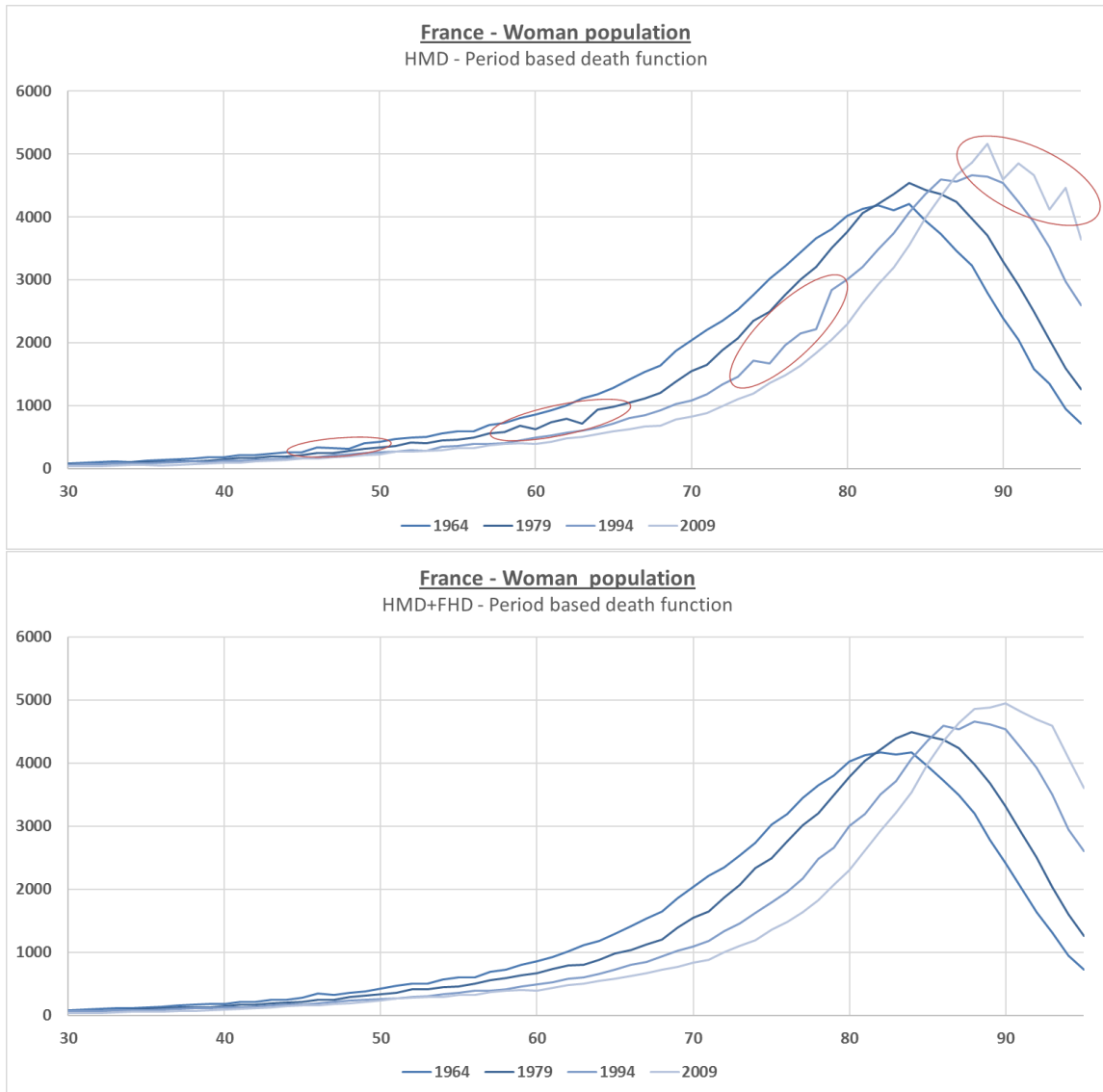

Figure 9 – Deaths curves, before and after HFD correction

In summary to the past section, analytical data shows that the correction added to HMD data effectively removes artificial cohort effects around generations 1918 to 1923, while keeping main mortality characteristics.

## 3. The impact of mortality data anomalies on an internal model

This section details the impact of retreated national population exposure on several classical mortality models. In a first step, we present the evolution of the fit of each mortality model used (M1, M3 and M5) and in a second step the impact on projected life expectancy based on both input sets (HMD versus HMD + HFD). Finally, we discuss how the retreated mortality data influence the selection of the best actuarial model, the computation of the Solvency Capital Requirement, as well as on the stability of the longevity risk assessment.

### 3.1. Internal model specification

The simplified prototype of internal model used for this analysis works as follows: a set of mortality models is fitted based on national population mortality data, and the 'best' model is selected based on both statistical and qualitative criteria. The selected mortality model is then used to draw simulated forecasts, and a prospective mortality table corresponding to the 99.5 percentile in terms of improvements is built. This allows to provide the longevity shock at the core of the Solvency II economic capital calculation, here based on an annuity product portfolio. Longevity risk can be divided in two different risk sources: a level component and a trend component, on which this article focuses on.

The internal model process workflow is illustrated in Figure 10: the usual workflow is represented in grey, and the additional step of mortality tables correction is represented in red. Indeed, this correction process is directly performed on national population mortality tables, and is then used in the model calibration and selection processes.

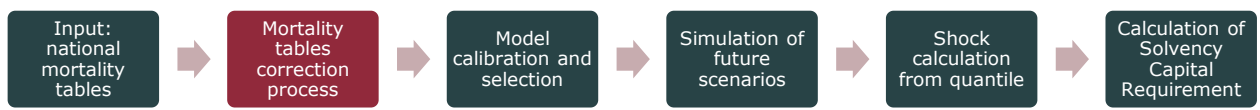

Figure 10 – The mortality table correction process within the Internal Model workflow

The simplified prototype of internal model, employed here to assess longevity risk, relies on several classical stochastic mortality models, which we briefly describe in the following.

The famous model by Lee and Carter (1992), also referred to as M1, decomposes mortality as a static age-structure $(\beta_x^{(1)})$, a general level driven by a stochastic process in time $(\kappa_t^{(2)})$ and an age-specific sensitivity $(\beta_x^{(2)})$ to this general level as follows:

$$log(m(t,x)) = \beta_x^{(1)} + \beta_x^{(2)}\kappa_t^{(2)}.$$

In order to account for possible cohort effects, while relying on a reasonable parametrization, one can also consider the Age-Period-Cohort model (M3) as proposed in Currie (2006), which is a special case of Renshaw & Haberman (2006), where the age, period and cohort components influence mortality independently in the following form:

$$log(m(t,x)) = \beta_x^{(1)} + \kappa_t^{(2)} + \gamma_{t-x}^{(3)},$$

with $\gamma_c$ the additional cohort-related factor which allows to adjust mortality rates for a 'generation' (or diagonal) which originates from year $c = t - x$.

As an alternative taking advantage of the log-linearity of mortality rates from intermediate ages, the CBD model from Cairns et al. (2006), known as M5, can also be tested:

$$logit(q(x,t)) = \kappa_t^{(1)} + \kappa_t^{(2)}(x - \bar{x}),$$

where $\bar{x}$ denotes the mean age over the range of ages used in the calibration.

To select the most adequate modelling to one set of historical mortality data, each model is calibrated independently in a first phase. Calibration results are compared in a second phase, putting emphasis on a set of necessary or wishful properties, for example:

- Quality of fit (BIC criterion) of historical mortality tables,
- Randomness of standardized residuals,
- Stability of model parameters to the age-band and historical-length of mortality data used for calibration,
- Back-test of projected mortality improvements against historical improvements.

For further discussion regarding interesting properties and tests for model selection, the reader is referred to e.g. Cairns et al. (2009). Calibration results will be discussed in the next section.

Let us recall that the data used to specify the crude death rates $m(x,t)$ by single year and one-year age bands were downloaded from the Human Mortality Database on September 1$^{st}$ 2015 for France (civil population), Italy (civil population) and West Germany (total population). As for calibration purposes, the datasets are restricted to the most recent decades of historic, retaining an age-band characteristic for a standard insured population. Mortality at higher ages than the maximum of the age band is commonly assessed based on a table closure technique. References and discussions on table closure methodologies can be found in Quashie and Denuit (2005).

## 3.2. Model fitting

For the purpose of illustration, we restrict our graphical analysis to France only, as results are very similar for Italy and Germany. Any specific outcome for Italy or Germany will be mentioned.

**M1 model**

The M1 model solely captures age-related and period related mortality effects. As the retreated exposures mostly correct cohort based effects, one could expect limited impacts on the M1 model components. As a matter of fact, differences of fit results on separate model components are very moderate, as illustrated below for the French female population.

Results shown on Figure 11 are similar for both genders and all countries in our study.

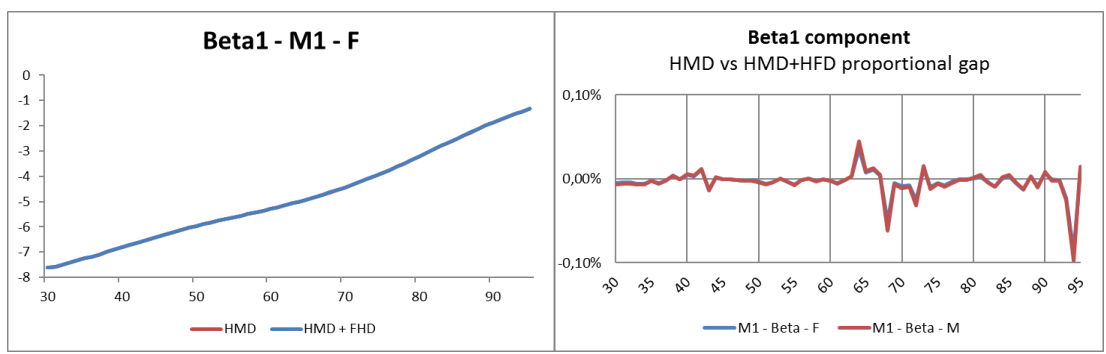

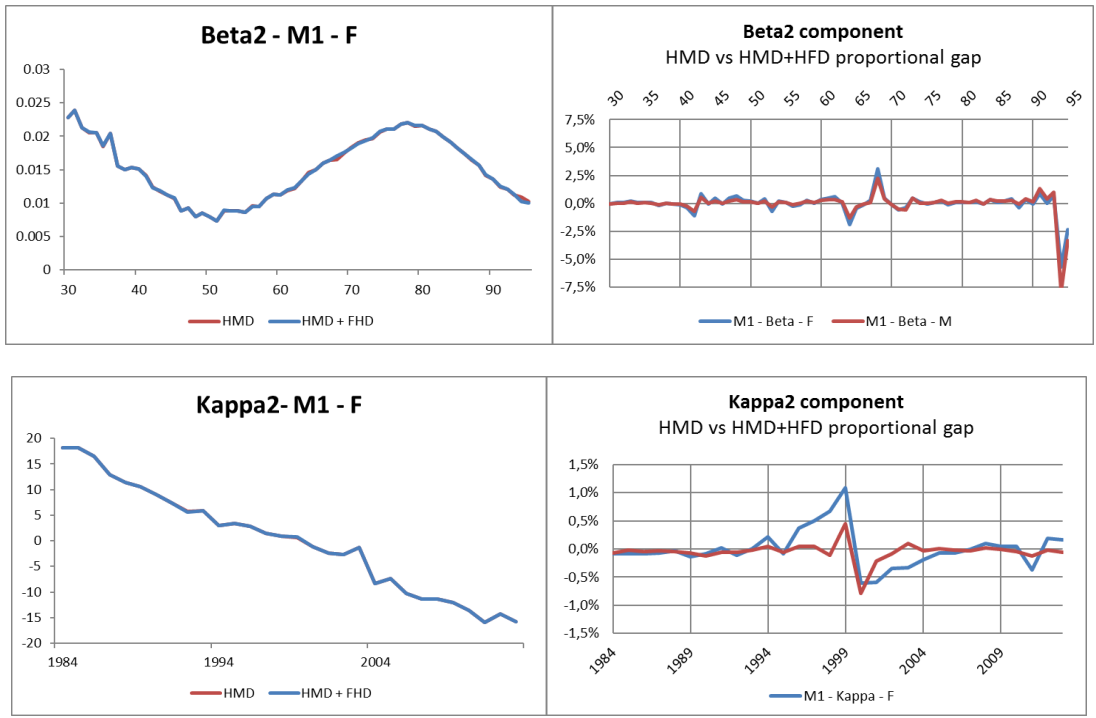

Figure 11 – Lee-Carter model (M1) components fitting

**M3 model**

In addition to age-related (beta) and period related (kappa) mortality effects, the M3 model also captures specific cohort effects between generations (gamma). We observe that the gamma component shows an improved stability on the specific generations related to shocks in birth patterns around the two World Wars, whereas the beta component (age-related) and kappa component (period related) show no sensible changes. Changes in the separate model components are restricted to specific generations of the cohort-effect component only.

Female and Male populations of each country in our study show similar results to the French female population illustrated below in Figure 12.

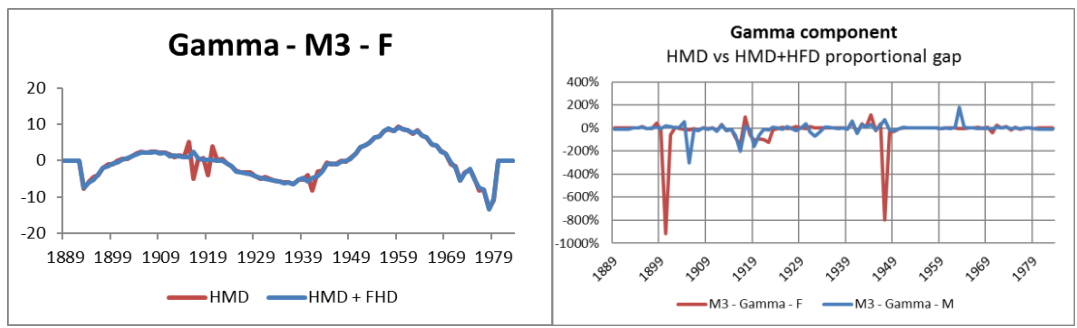

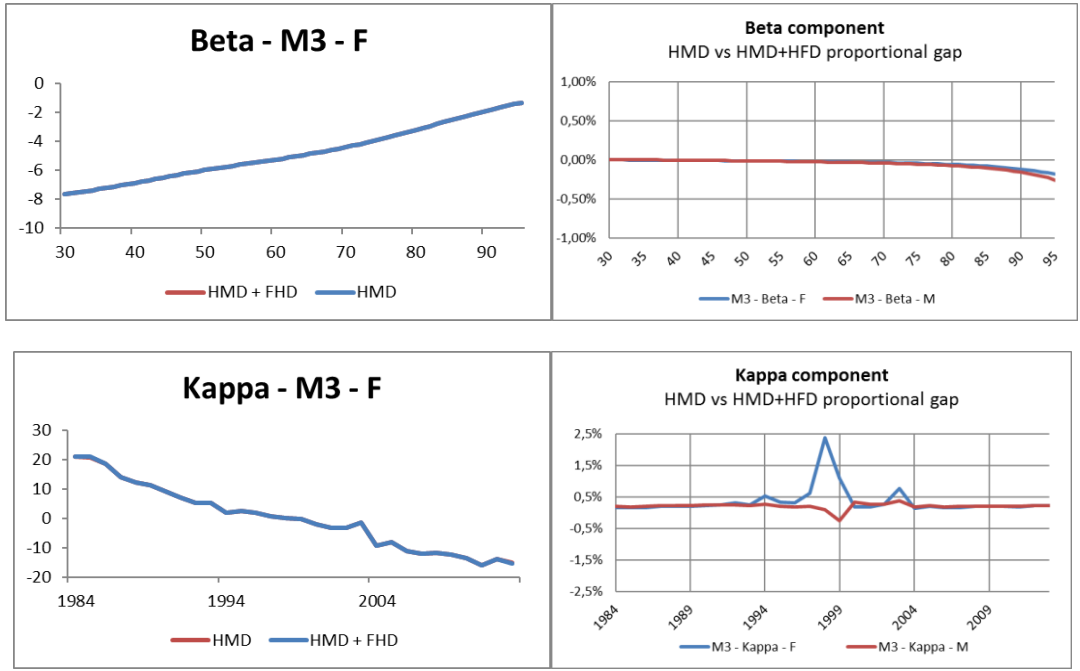

Figure 12 – Age-Period-Cohort model (M3) components fitting

**M5 model**

The M5 model solely captures age related and period related mortality effects. In our test, it proved to be very stable against the new retreated HMD + HFD tables for both gender in each country of our study (see Figure 13).

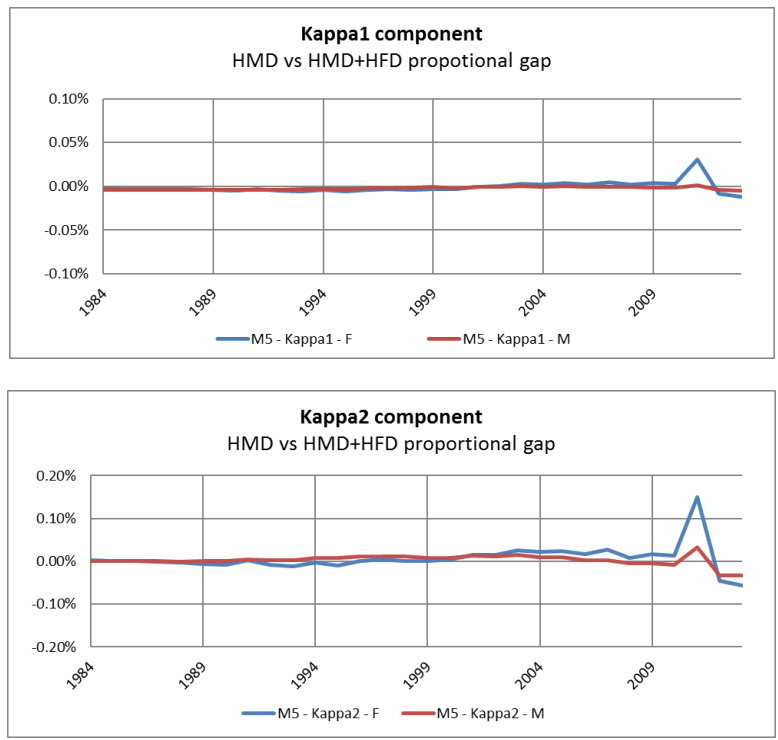

Figure 13 – Cairns-Blake-Dowd model (M5) components fitting

## 3.3. Impact on future Life expectancy

In this section, we propose to highlight potential changes on Life expectancy through two complementary approaches. On the one hand, projected period mortality tables are compared with and without exposure retreatment on the basis of residual period life expectancy at age 30 and truncated at age 95. On the other hand, cohort mortality tables are compared with and without exposure retreatment for all active generations being of age 30 to 95 years in 2015. Figures present the outcome of the models stochastic projection and differences between life expectancies for the median (baseline), 0.5$^{th}$ percentile (mortality shock) and 99.5$^{th}$ percentile (longevity shock) scenarios.

**M1 model**

As it was expected based on the reduced impact on the fit of M1 model component, Life expectancy analysis show very limited change between both set of population's exposure data, as depicted in Figure 14. For both genders and for each country in our study, period-based life expectancy between age 30 and 95 as well as live generation life expectancy do not vary more than half a month.

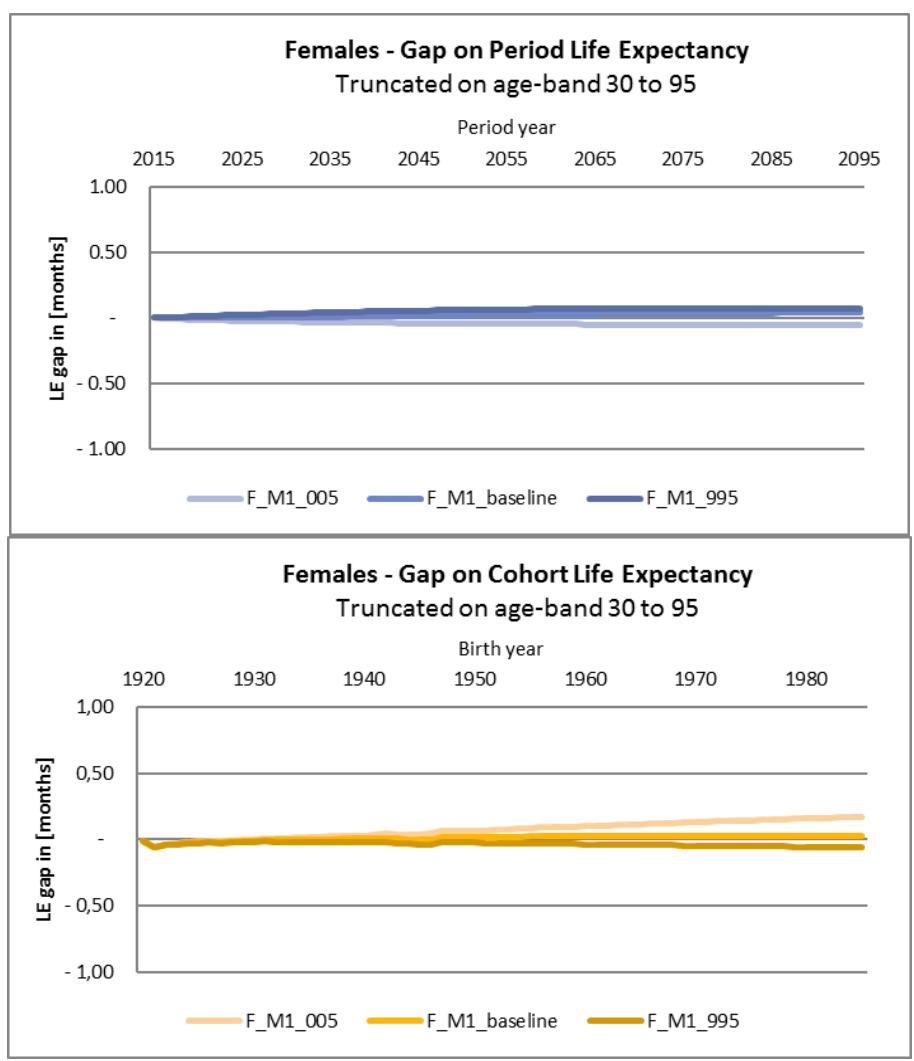

Figure 14 – Period and Cohort life expectancy as of 2015 for French females, M1 model

### M3 model

Detailing the impact on the M3 model calibration, we observe that the projected life expectancy (period-based) is very stable. Cohort Life expectancy show limited variations, with highest impact located during World War II.

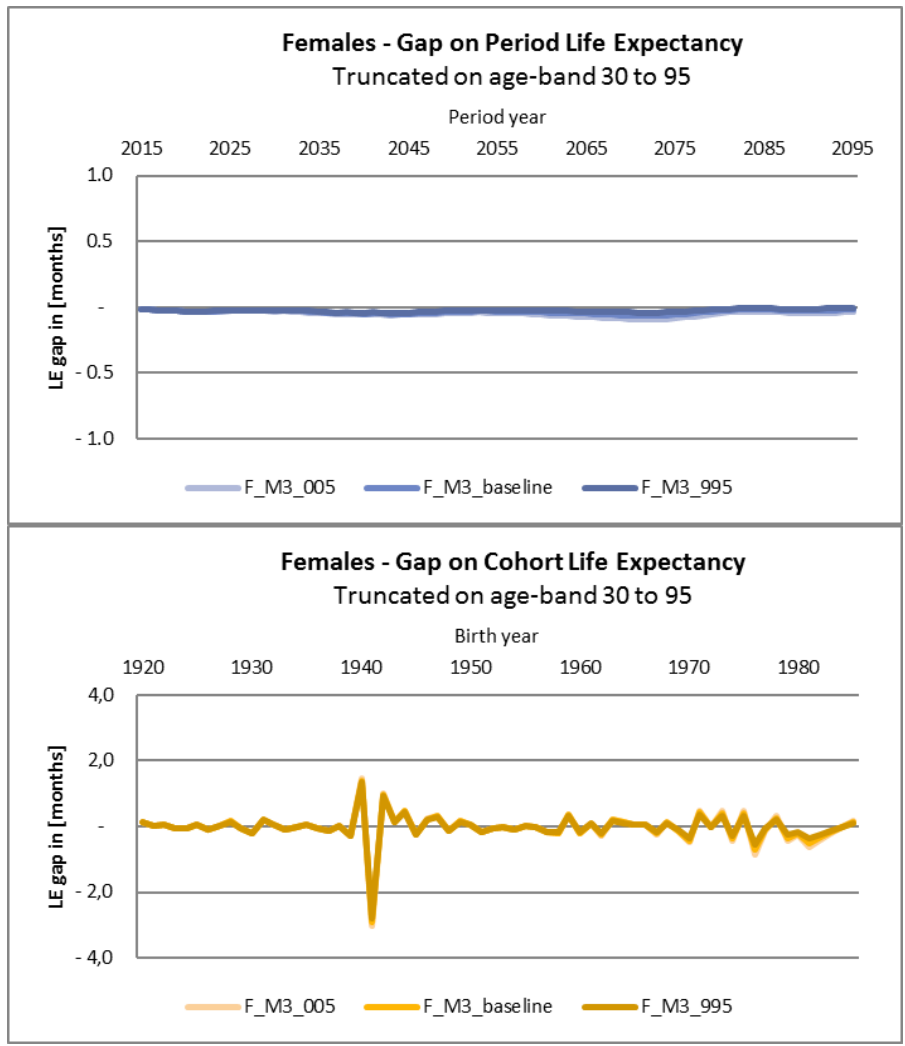

**Figure 15 – Period and Cohort life expectancy as of 2015 for French females, M3 model**

### M5 model

Turning to the M5 model, as the stability between the fits of the model components led to think, the retreated exposure data have no material impact on period life expectancy and on cohort life expectancy (see Figure 16).

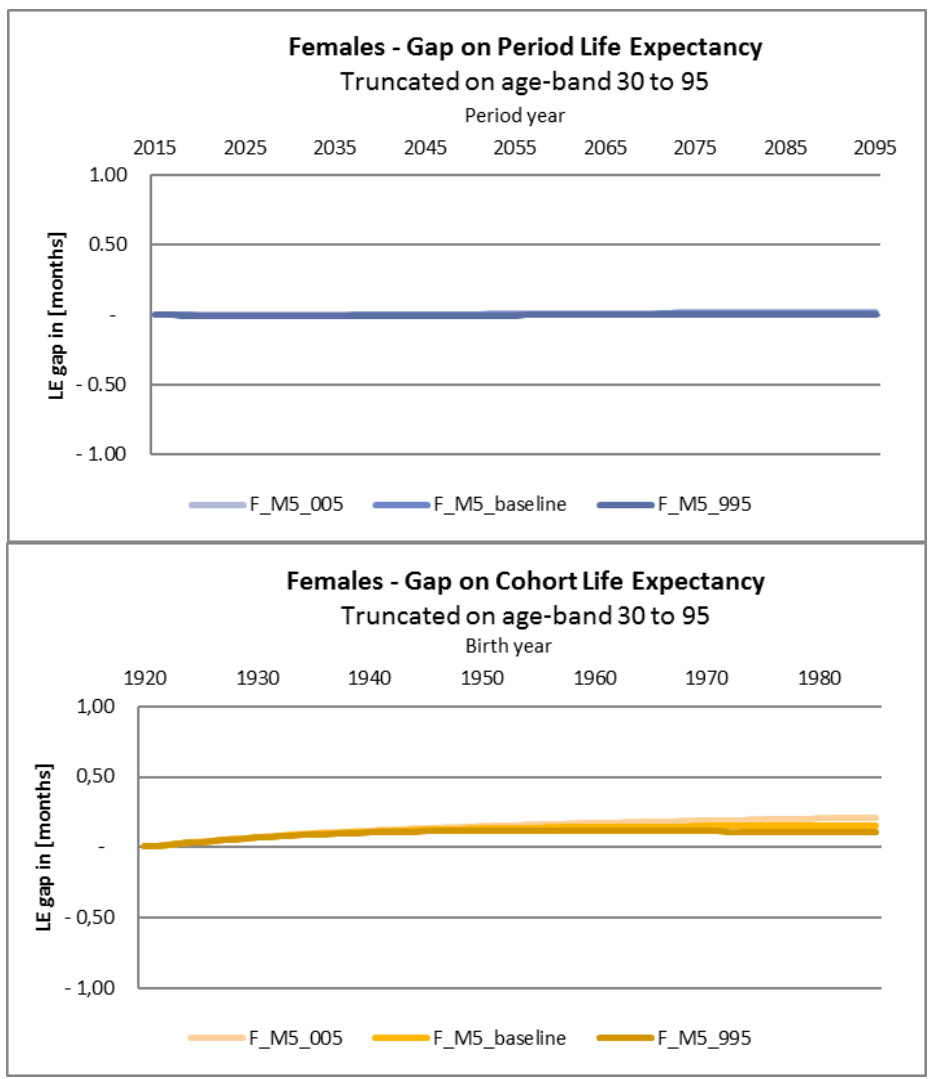

Figure 16 – Period and Cohort life expectancy as of 2015 for French females, M5 model

**Conclusion**

In a nutshell, the study of the model fits on historical tables shows that M1 and M5 models are almost not impacted by the cohort-effect retreatment, due to the models not capturing the cohort effect. M3 model is not significantly impacted by the corrected exposure tables as both period and cohort-based life expectancy keep the same level. However, cohort Life expectancy shows evidences of change in the inter-generational structure, and the volatility is reduced in the model calibrated with HMD with HFD correction (see again Figure 12).

## 3.4. Model selection process

As far as the model selection process is concerned, in all cases of our study, the improved BIC criterion as depicted in Figure 17 shows that the retreated exposure allows M1 and M5 model to better capture mortality dynamics embedded in HMD table. BIC criterion does not show a measurable change for the M3. Finally, for several populations as for Italian female table, retreated exposures allow for M1 to achieve a higher BIC level than the M3 model.

As highlighted in the former section, the retreatment of the exposure reduces fluctuations of the component related to the cohort-effect. Therefore, models that do not take cohort-effect into account improve their efficiency to fit historical mortality data; models that capture cohort-effect also benefit from an improved stability (less fluctuation).

| France Females | BIC criterion | | |
|---|---|---|---|
| | HMD | HMD+HFD | Abs. diff. |
| M1 | -13 585 | -11 907 | -12% |
| M3 | -12 625 | -12 625 | 0% |
| M5 | -160 771 | -159 199 | -1% |

| France Males | BIC criterion | | |
|---|---|---|---|
| | HMD | HMD+HFD | Abs. diff. |
| M1 | -16 767 | -15 128 | -10% |
| M3 | -13 627 | -13 627 | 0% |
| M5 | -72 421 | -70 805 | -2% |

| Italy Females | BIC criterion | | |
|---|---|---|---|
| | HMD | HMD+HFD | % diff. |
| M1 | -14 217 | -12 427 | -13% |
| M3 | -13 596 | -13 596 | 0% |
| M5 | -65 430 | -63 734 | -3% |

| Italy Males | BIC criterion | | |
|---|---|---|---|
| | HMD | HMD+HFD | % diff. |
| M1 | -18 117 | -16 216 | -10% |
| M3 | -15 026 | -15 026 | 0% |
| M5 | -29 882 | -27 939 | -7% |

| Germany Females | BIC criterion | | |
|---|---|---|---|
| | HMD | HMD+HFD | % diff. |
| M1 | -16 410 | -13 360 | -19% |
| M3 | -13 478 | -13 478 | 0% |
| M5 | -99 971 | -96 901 | -3% |

| Germany Males | BIC criterion | | |
|---|---|---|---|
| | HMD | HMD+HFD | % diff. |
| M1 | -16 363 | -14 318 | -13% |
| M3 | -12 959 | -12 959 | 0% |
| M5 | -28 060 | -25 971 | -7% |

Figure 17 – BIC criteria for model calibration

## 3.5. Solvency Capital Requirement

The trend component of the longevity SCR (called "SCR" in the following paragraphs) is here given as a change of future mortality improvements. The model described previously gives a full distribution of mortality improvement rate by gender, age and year. The most adverse scenario corresponds to the 0.5$^{th}$ percentile of the distribution of longevity improvements, denoted $IR^{SCR}_{x,t_0 \to t}$. To obtain the final mortality table for SCR calculations, the improvements of the Best Estimate assumptions

$$IR^{BE}_{x,t_0 \to t} = \frac{q^{BE}_{x,t} - q^{BE}_{x,t_0}}{q^{BE}_{x,t_0}}, for\ t > t_0,$$

are replaced by the ones obtained from the model mentioned above for all years from the valuation year $t_0$ onwards. The initial mortality rates of the trend SCR component for year $t_0$ do not change from the Best Estimate mortality rates for this same year. For future years $t > t_0$, the mortality rates are defined as follows for the Best Estimate ($q^{BE}_{x,t}$) and the SCR ($q^{SCR}_{x,t}$) respectively:

$$q^{BE}_{x,t} = q^{BE}_{x,t_0} \cdot \left(1 + IR^{BE}_{x,t_0 \to t}\right),$$

$$q^{SCR}_{x,t} = q^{BE}_{x,t_0} \cdot \left(1 + IR^{SCR}_{x,t_0 \to t}\right).$$

We depict on Figure 18 the impact on mortality rate scenarios of the change of data source (HMD data with HFD correction). As the improvements distribution is changed with the data correction, we observe on Figure 18 a slight impact on mortality rates due to lower cohort effects, as stated previously during the calibration of mortality models. Note that both the baseline scenario ("Base") and the SCR scenario are changed in this figure.

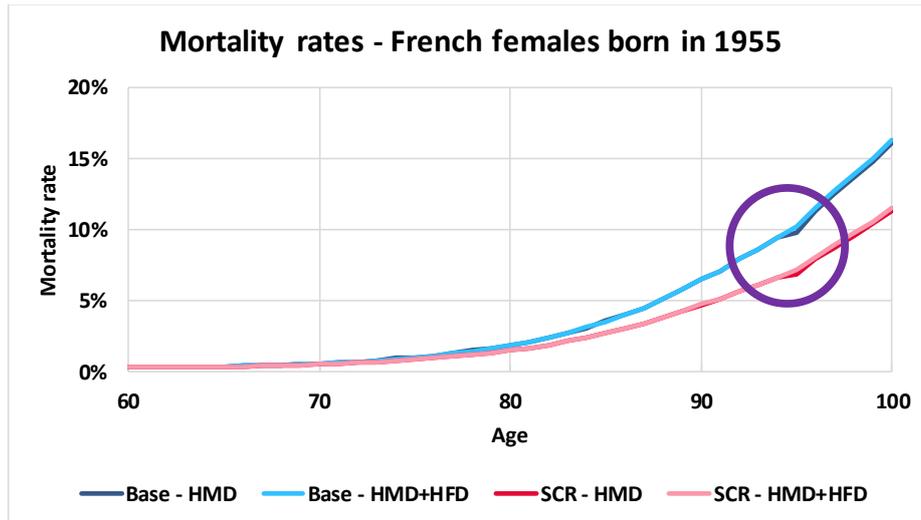

Figure 18 – Baseline and Longevity Trend SCR future mortality for French females born in 1955

As stated previously, there is less volatility in mortality assumptions when using the HFD correction. The level of mortality rates is however not impacted. The Best Estimate used are already without this excess of volatility, therefore in the following analysis we will focus only at the impact on SCR from changing the HMD data with HFD-corrected data while keeping the same Best Estimate assumptions.

This decrease of captured volatility in the SCR scenario can be seen for various different cohorts. This can be illustrated with the impact of Longevity Trend SCR on the evolution of the cohort life expectancy. The cohort life expectancy including future mortality improvements is defined as follows for $A \in \{BE, SCR\}$:

$$e_{x,t}^{Cohort}(A) = \sum_{k=1}^{\infty} \prod_{i=0}^{k-1} (1 - q_{x+i,t+i}^{A}).$$

The impact on cohort life expectancy from replacing the Best Estimate improvements by the SCR improvements is defined as follows:

$$IE_{x,t} = \frac{e_{x,t}^{Cohort}(SCR) - e_{x,t}^{Cohort}(BE)}{e_{x,t}^{Cohort}(BE)}.$$

The impact is illustrated on Figure 19 for French females in 2015 born between 1925 and 1975, when having on one hand the Best Estimate assumptions with the improvement remaining the same, and on the other mortality rates using Longevity SCR improvements. For the later, we perform the calculations with the data from HMD only, and with the HMD data corrected with the HFD data.

With the change of data we observe a lower volatility on the indicator $IE_{x,t}$ for some cohorts, notably females being born in the early 1940's during World War II, i.e. being between 70 and 75 in 2015. There is a small difference for older people (those being born between the World Wars) and almost no difference for those being born after 1945.

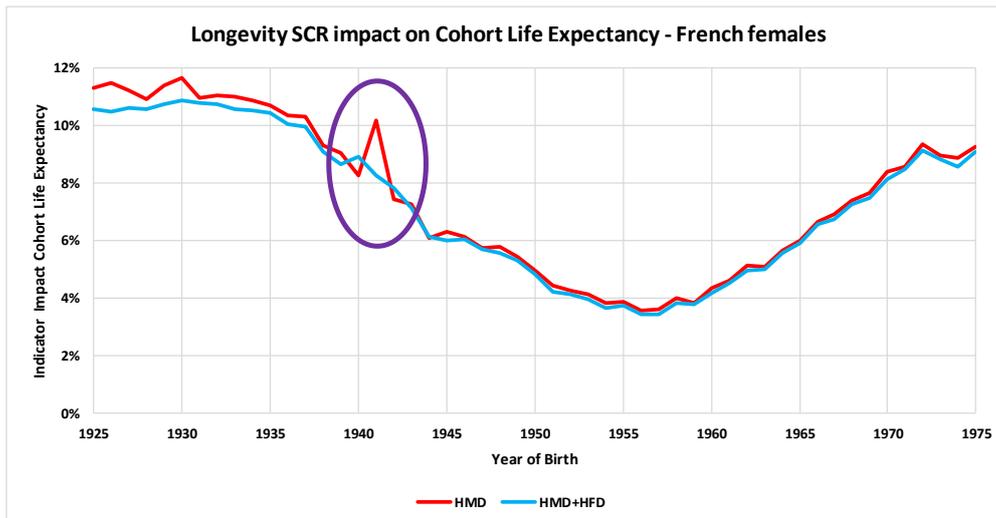

**Figure 19 – Longevity SCR impact on Cohort Life Expectancy in 2015 for French females born between 1925 and 1975**

In order to compute the final impact, the portfolios used for the study are originated from the three countries France, Germany and Italy, including various products. For France, it is a savings and retirement business made of individual and group products, with on the in-force an average age around 65/70 for the annuity phase. For Germany these are individual investment & savings products, with an average age of policyholders around 47. For Italy the products are investment & savings products (62%) and long-term care products (38%) exposed to longevity risk. The savings and retirement businesses for these three countries are a mix of General Account, with various guaranteed rates, and Unit-Linked products for the accumulation phase. Long-term care are a pure General Account products.

Using the full Internal Model methodology and the aggregation with other risks, with on one hand the HMD data and on the other hand the HMD data corrected with HFD, we observe that the correction leads to a small reduction of Longevity SCR (Figure 20).

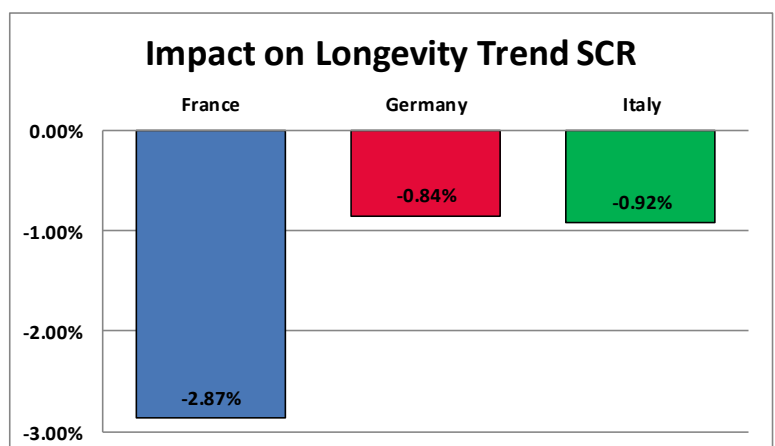

**Figure 20 – Impact on Longevity Trend SCR**

This negative impact is consistent with the results found previously on indicator $IE_{x,t}$: the removal of non-necessary volatility on several cohorts lead to a decrease of the Longevity SCR. This is even clearer for the French portfolio, when compared to the German and Italian portfolios, as the French portfolio contains more annuitants close to ages 70-75 at which the reduction of non-necessary volatility is the strongest.

Looking now at the total Life SCR of each of these entities with the other risks and the diversification between these risks calculated with an Internal Model methodology, the impacts on total Life SCR are negative but almost non-significant (Figure 21) even as Longevity STEC roughly represents 20% Life STEC pre-diversification.

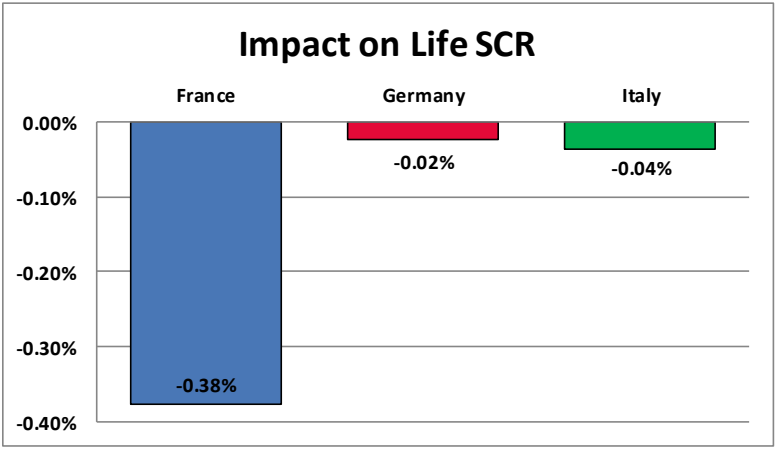

Figure 21 – Impact on Life SCR

Even if the quantitative impacts are very moderate, the treatment of HMD data with the HFD database has a positive effect on the data quality and therefore on the longevity risk assessment due to a lower volatility in future mortality assumptions.

## 3.6. Stability of the longevity risk assessment

The stability from a year to another of the longevity risk assessment can be roughly estimated by looking at the evolution of the impact of Longevity Trend SCR mortality improvements on the cohort life expectancy of the portfolio, *i.e.* the difference for a policyholder at age $x$ in year $t$:

$$e_{x,t}^{Cohort}(SCR) - e_{x,t}^{Cohort}(BE)$$

and its evolution between t and t+1. This indicator is calculated on the whole portfolio mix. There again, the Best Estimate does not change but SCR mortality improvements change as we consider the improvements calibrated with HMD data, and also the improvements calibrated with HMD retreated with HFD data.

As shown in Figure 22, the retreatment brings a more stable longevity risk assessment between 2015 and 2016 for France and Italy but not for Germany, notably due to women. Overall, the volatility of Longevity SCR shock over time decreases.

| Evolution with SCR scenario from | France | Germany | Italy |
|---|---|---|---|
| HMD data | -2.31% | 0.40% | -1.01% |
| HMD+HFD data | 0.26% | 1.35% | 0.06% |

Figure 22 – Evolution of Cohort Life Expectancy gap between baseline and Longevity SCR Trend between 2015 and 2016

## 4. Concluding remarks

To illustrate it with a caricature, even if an internal model is the most robust one from a technical standpoint, is perfectly adapted to the inherent risks of a company, the "garbage in, garbage out" effect has to be avoided. In a global context of increasing regulatory needs, particularly on the data quality, using such a methodology to improve the data of external providers, as it is the case for the Human Mortality Database, is key. This database is regarded as a reference on the insurance market to have national mortality data on a large scope of countries, with similar demographic methodologies used from one country to another. Implementing the described methodology to correct data anomalies can therefore help the insurance market.

Risks are consequently better captured, assessed and monitored. The conclusion of the study and the practical application show that the longevity assessment remains stable as the global impact is non-significant on the tested portfolios but indicators are improved; with a better data quality the volatility coming from these exceptional cohort, effects is lowered, here slightly decreasing the capital requirement.

This kind of database cleansing is obviously welcomed and can help demographers and actuaries to better monitor longevity and mortality risks. In the future, it is important to continue tracking and correcting other kinds of errors that can be included in the current mortality tables.